\documentclass[12pt,twoside]{article}
\usepackage{fleqn,espcrc1}

\usepackage{graphicx}
\usepackage[figuresright]{rotating}

\newcommand{\AmS}{{\protect\the\textfont2
  A\kern-.1667em\lower.5ex\hbox{M}\kern-.125emS}}

\title{Fluctuation probes of quark deconfinement}

\author{M. Asakawa\address[MA]{Department of Physics, Nagoya University,
Nagoya 464-8602, Japan},
       U. Heinz\address[UH]{Department of Physics, The Ohio State University,
Columbus, OH 43210, U.S.A.}
        and
       B. M\"{u}ller\address[BM]{Department of Physics, Duke University,
Durham, NC 27708-0305, U.S.A.}}

\begin{document}

\maketitle

\vspace*{1cm}

Fluctuations in the multiplicities and momen\-tum distributions of 
particles emitted in relativistic hea\-vy-ion collisions have been 
widely considered as probes of thermalization and the statistical 
nature of particle production in such reactions \cite{Mrow}. The 
characte\-ri\-stic behavior of temperature and pion multiplicity 
fluctuations in the final state has been proposed as a tool for the
measurement of the specific heat \cite{Sto95} and, specifically, for 
the detection of a critical point in the nuclear matter phase 
diagram \cite{SRS}.

Two recent papers \cite{ahm00,jk00} drew attention to a different 
type of fluctuations, which is sensitive to the microscopic structure 
of the dense matter. If the expansion is too fast for local fluctuations 
to follow the mean thermodynamic evolution of the system, it makes 
sense to consider fluctuations of locally conserved quantities that 
show a distinctly different behavior in a hadron gas (HG) and a 
quark-gluon plasma (QGP). Characteristic features of the plasma 
phase may then survive in the finally observed fluctuations. In this
spirit we here focus on fluctuations of the net baryon number and
net electric charge as probes of the transition from hadronic matter 
to a deconfined QGP. We consider matter which is meson-dominated, which
renders our arguments applicable to SPS energies and above. 

In a hadron gas nearly two thirds of the hadrons (for $\mu\ll T$ mostly 
pions, where $\mu$ and $T$ are baryonic chemical potential and temperature, 
respectively) carry electric charge $\pm 1$. In the deconfined
QGP phase the charged quarks and antiquarks make up only about half 
the degrees of freedom, with charges of only $\pm {1\over 3}$ or 
$\pm {2\over 3}$. Consequently, the fluctuation of one charged 
particle in or out of the considered subvolume produces a larger 
mean square fluctuation of the net electric charge if the system is 
in the HG phase. For baryon number fluctuations the situation is 
less obvious because in the HG baryon charge is now only carried by 
the heavy and less abundant baryons and antibaryons. Still, all of 
them carry unit baryon charge $\pm 1$, while the quarks and antiquarks 
in the QGP only have baryon number $\pm {1\over 3}$. It turns out 
that, as $\mu/T \to 0$, the fluctuations are again larger in the HG, 
albeit by a smaller margin than for charge fluctuations. At SPS 
energies and below the difference between the two phases increases 
since the stopped net baryons contribute to the fluctuations, and 
more so in the HG than in the QGP phase.   
  
The value of a conserved quantum number of an isolated system does not 
fluctuate at all. However, if we consider a small part of the system, 
which is large enough to neglect quantum fluctuations, but small 
enough that the entire system can be treated as a heat bath, 
the statistical uncertainty of the value of the observable in the 
subsystem can be calculated. This is the scenario considered here.

We first discuss fluctuations of the net baryon number. In the dilute 
HG phase, we can apply the Boltzmann approximation. The net baryon 
number is $N_b\,{=}\,N_b^{+}{-}N_b^{-}$, where $N_b^{\pm}$ denotes the 
number of baryons $(+)$ and antibaryons $(-)$, respectively. Then the 
net baryon number fluctuations in the hadronic gas are given by
 \begin{equation}
    (\Delta N_b)_{\rm HG}^2 = 2\, N_b (T)\, \cosh(\mu/T)\, . 
 \label{eq4}
 \end{equation}
The late expansion of the fireball being nearly isentropic, the ratio 
of baryon number fluctuations to entropy $S$, $(\Delta N_b)^2/S$,
provides a useful measure for the early fluctuations. For a transient 
quark phase the baryon fluctuations are given by \cite{fn1} 
 \begin{equation}
   \left. {(\Delta N_b)^2 \over S}\right\vert_{\rm QGP} 
   = {5\over 37\pi^2}
   \left( 1 + {22\over 111}\Bigl({\mu\over \pi T}\Bigr)^2 + \ldots
   \right) \, ,
 \label{eq7}
 \end{equation}
assuming an ideal gas of massless quarks and gluons with two massless
flavors. The entropy can be estima\-ted from the final hadron 
multiplicity \cite{SH92}. For high collision energies ($\mu/T{\,\to\,}0$), 
the ratio (\ref{eq7}) ap\-proaches a constant; even for SPS energies, 
the $\mu$-de\-pen\-dent correction is at most 5\%. The many resonance 
contributions make it difficult to write down an analytic expression
for $S_{\rm HG}/V$, but it is clear from (\ref{eq4}) that the 
$\mu$-dependence of the corresponding ratio in the HG phase is stronger 
than for the QGP phase. This translates into a stronger beam energy 
dependence of (\ref{eq7}) near midrapidity.

The results for net charge fluctuations are similar. All stable charged 
hadrons have unit electric charge; again using the Boltzmann approximation
we find $(\Delta Q)_{\rm HG}^2 = N_{\rm ch}$,
where $N_{\rm ch}$ is the total number of charged particles emitted
from the subvolume. The ratio $(\Delta Q)^2/S$ for the QGP is a factor 
${5\over 2}$ larger than the corresponding ratio (\ref{eq7}) for baryon 
number fluctuations, due to the larger electric charge of the $u$ quarks, 
but shows the same weak $\mu$-dependence. The main difference to baryon 
number fluctuations arises in the HG phase: Since at SPS and higher 
energies $N_{\rm ch}$ is dominated by pions and meson resonances, 
its $\mu$-dependence is now also weak. In contrast to baryon 
number fluctuations, charge fluctuations thus show a weak beam energy 
dependence in either phase, and only their absolute values differ.

Numerical values for the ratios $(\Delta Q)^2/S$ and $(\Delta N_b)^2/S$
at SPS energies and above were derived in \cite{ahm00} and are shown in 
Fig.~1. The fluctuations in the QGP are typically a factor 2 below those
in the HG, with a somewhat smaller reduction for baryon number than for 
charge fluctuations. 

\begin{figure}[tbh]
\begin{center}
\includegraphics[scale=0.5]{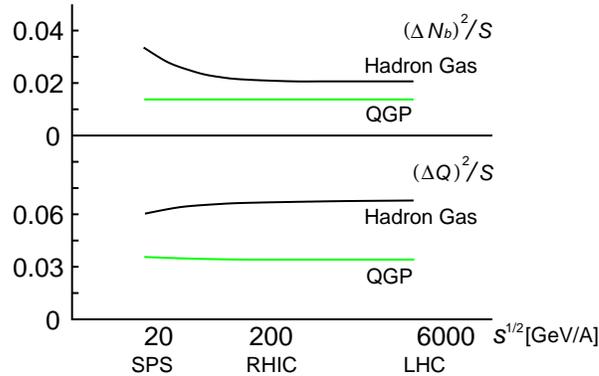}
\end{center}
\vspace*{-0.8cm}
\caption{Schematic drawing of the beam energy dependence of the net
  baryon number and charge fluctuations per unit entropy
  for a hadronic gas and a quark-gluon plasma. \label{fig1}}
\end{figure}

These estimates, including our corrections for resonance decays, refer 
to ideal gases in equilibrium. Future work should address interaction 
effects on the thermal fluctuations in HG and QGP \cite{jk00} and treat 
resonance decays kinetically. We also point out potentially important 
non-equilibrium aspects: The fluctuation/entropy ratios in the QGP 
will be even lower (facilitating the discrimination against HG) if 
initially the QGP is strongly gluon dominated \cite{Sh92} and thus 
may lie below the equilibrium value (1) if the QGP hadronizes 
before the concentrations of the (baryon) charge carriers $q$ and 
$\bar q$ saturate \cite{Betal}.

We now discuss whether the difference between the two phases is 
really observable. Even if a QGP is temporarily created in a heavy-ion 
collision, all hadrons are emitted after re-hadronization. Thus, it is 
natural to ask whether the fluctuations will not always reflect the 
hadronic nature of the emitting environment. It is essential to our 
argument that fluctuations of conserved quantum numbers can only be 
changed by particle transport (diffusion) and, due to the rapid expansion 
of the reaction zone, are likely to be frozen at an early stage. For our 
estimate we assume for simplicity that the fire\-ball expands mostly 
longitudinally, with a boost in\-va\-ri\-ant (Bjorken) flow profile. 
Strong longitudinal flow exists in collisions at the SPS \cite{NA49-HBT}, 
and the Bjorken picture is widely expected to hold for collisions at 
RHIC and LHC. 

We here give a qualitative argument for the survival of a baryon number 
fluctuation within a rapi\-di\-ty interval $\Delta\eta{\,\approx\,}1$ 
\cite{fn2}. At RHIC energies, the reaction zone takes about 9 fm/$c$ to 
cool from the hadronization temperature to the hadronic freeze-out point, 
and during this time this rapidity interval expands from a length of 
5\,fm to 14\,fm \cite{ahm00}. We consider the diffusion of net baryon 
number in and out of this rapidity interval by hadronic rescattering 
processes after QGP hadronization. Baryons have an average thermal 
longitudinal velocity component ${\bar v}_z{\,=\,}0.325$ at the 
hadronization temperature $T{\,\approx\,}170$ MeV. 
Without rescattering, a baryon which at the point of hadronization  
is at the center of this interval can travel on average only about 3\,fm 
in the beam direction before freeze-out. Hence it will not reach the edge 
of the rapidity interval. Rescattering in the hot hadronic matter inhibits
baryon number diffusion, and a fluctuation will even survive in a smaller 
rapidity interval \cite{ahm00}. We conclude that the short time 
between hadronization and final freeze-out precludes the readjustment of 
net baryon number fluctuations in rapidity bins $\Delta\eta\ge 1$. A similar 
calculation applies to net charge fluctuations.

Stephanov and Shuryak recently derived a diffusive transport equation
describing the evolution of fluctuations of conserved quantum numbers
during the hydrodynamic expansion of a dense fireball \cite{foot3}. Within 
this general framework they argued that quasi-elastic particle reactions 
through resonances such as $\pi\pi\rightarrow\rho\rightarrow\pi\pi$ and 
$\pi N\rightarrow\Delta\rightarrow\pi N$ lead to substantial
rapidity shifts for the involved particles (in particular the pions)
and thus dominate net baryon and electric charge diffusion in the 
hadronic phase. However, with realistic estimates for the number
of such reactions between hadronization and freeze-out, even the more 
fragile charge fluctuation signal should still survive hadronic 
rescattering in rapidity windows $\Delta \eta > 3$ \cite{KBJ01}.
The average rapidity shift for a nucleon is a factor four smaller 
than for pions, such that the critical rapidity window size for the 
survival of net baryon number fluctuations is less than unity, in 
agreement with our estimate \cite{ahm00}. 

Due to local charge and baryon number conservation, the rapid 
longitudinal expansion of the reaction zone may even freeze the 
fluctuations established during the initial particle production process 
\cite{gm00}, before reaching the equilibrium level corresponding to a 
thermalized QGP (even if average thermodynamic quantities reach their
thermal values). A detailed study of initial state fluctuations would 
thus be desirable.

In conclusion, we have argued that the difference in magnitude of 
local fluctuations of the net baryon number and net electric charge 
between confined and deconfined hadronic matter is partially frozen
at an early stage in relativistic heavy-ion collisions. These
fluctuations may thus be useful probes of the temporary formation of 
a deconfined state in such collisions. The event-by-event 
fluctuations of the two suggested observables for collisions with a 
fixed value of the transverse energy $dE_{\rm T}/dy$ or of the 
energy measured in a zero-degree calorimeter would be appropriate
observables that could test our predictions.



\begin{thebibliography}{99}
\bibitem{fn1}
  This formula ignores strange particles, but in the results shown below
  they are properly included \cite{ahm00}.
\end{thebibliography}
\end{document}